\documentclass{kapedbk} 
\usepackage{edbkps}     
\usepackage{times}     
\usepackage{epsfig,rotating}      
\usepackage{graphicx,dcolumn,xspace}
\usepackage{amsmath}
\usepackage{amsfonts}
\usepackage{amssymb}
\newcolumntype{.}{D{.}{.}{1}}

\sectionequationnumber 
\upperandlowercase
\chapsectrunningheads
\chaptersection

\begin{document}
\newcommand\adsr{{Adv.~Space~Res.}}%
\newcommand\jatp{{J.~Atmos.~Terres.~Phys.}}%
\newcommand\aj{{Astron.~J.}}%
\newcommand\actaa{{Acta Astron.}}%
\newcommand\areps{{Ann.~Rev.~Earth \& Plan.~Sci.}}%
\newcommand\araa{{Ann.~Rev.~Astron.~\& Astrophys.}}%
\newcommand\apj{{Astrophys.~J.}}%
\newcommand\apjl{{Astrophys.~J.~Let.}}%
\newcommand\apjs{{Astrophys.~J.~Supl.}}%
\newcommand\ao{{Appl.~Opt.}}%
\newcommand\apss{{Astrophys.~\& Space Sci.}}%
\newcommand\aap{{Astron.~\& Astrophys.}}%
\newcommand\aapr{{Astron.~ \& Astrophys.~Rev.}}%
\newcommand\aaps{{Astron.~ \& Astrophys. Supl.}}%
\newcommand\azh{{Astron.~Zh.}}%
\newcommand\baas{{Bull.~Amer.~Astron.~Soc.}}%
\newcommand\caa{{Chinese Astron. Astrophys.}}%
\newcommand\cjaa{{Chinese J. Astron. Astrophys.}}%
\newcommand\icarus{{Icarus}}%
\newcommand\jcap{{J. Cosmology Astropart. Phys.}}%
\newcommand\jrasc{{JRASC}}%
\newcommand\memras{{Mm.~Roy.~Astron.~Soc.}}%
\newcommand\mnras{{Mon.~Not.~ Roy.~Astron.~Soc.}}%
\newcommand\na{{New Astron.}}%
\newcommand\nar{{New Astron.~Rev.}}%
\newcommand\pra{{Phys.~Rev.~A}}%
\newcommand\prb{{Phys.~Rev.~B}}%
\newcommand\prc{{Phys.~Rev.~C}}%
\newcommand\prd{{Phys.~Rev.~D}}%
\newcommand\pre{{Phys.~Rev.~E}}%
\newcommand\prl{{Phys.~Rev.~Lett.}}%
\newcommand\pasa{{PASA}}%
\newcommand\pasp{{Pub.~Astron.~Soc.~Pac.}}%
\newcommand\pasj{{PASJ}}%
\newcommand\qjras{{QJRAS}}%
\newcommand\rmxaa{{Rev. Mexicana Astron. Astrofis.}}%
\newcommand\skytel{{S\&T}}%
\newcommand\solphys{{Sol.~Phys.}}%
\newcommand\sovast{{Soviet~Ast.}}%
\newcommand\ssr{{Space~Sci.~Rev.}}%
\newcommand\zap{{Zeit.~Astrophy.}}%
\newcommand\nat{{Nature}}%
\newcommand\iaucirc{{IAU~Circ.}}%
\newcommand\aplett{{Astrophys.~Lett.}}%
\newcommand\apspr{{Astrophys.~Space~Phys.~Res.}}%
\newcommand\bain{{Bull.~Astron.~Inst.~Netherlands}}%
\newcommand\fcp{{Fund.~Cosmic~Phys.}}%
\newcommand\gca{{Geochim.~Cosmochim.~Acta}}%
\newcommand\grl{{Geophys.~Res. ~Lett.}}%
\newcommand\jcp{{J.~Chem.~Phys.}}%
\newcommand\jgr{{J.~Geophys.~Res.}}%
\newcommand\jqsrt{{J.~Quant.~Spec.~Radiat.~Transf.}}%
\newcommand\memsai{{Mem.~Soc.~Astron.~Italiana}}%
\newcommand\nphysa{{Nucl.~Phys.~A}}%
\newcommand\physrep{{Phys.~Rep.}}%
\newcommand\physscr{{Phys.~Scr}}%
\newcommand\planss{{Planet.~Space~Sci.}}%
\newcommand\procspie{{Proc.~SPIE}}%


\def\ebv{$E$(B-V)}
\def\glong{$\ell$}
\def\nHI{\hbox{$n$(H$^\mathrm {o }$)}}
\def\nHII{\hbox{$n$(H$^\mathrm {+ }$)}}
\def\Beten{\hbox{$^{ 10 }$Be}}
\def\Cfourteen{\hbox{$^{ 14 }$C}} 
\def\tauhalf{\hbox{$\tau _ {1/2}$}}
\def\Fesixty{\hbox{$^{ 60 }$Fe}}
\def\Kforty{\hbox{$^{ 40 }$K}}
\def\Clthirtysix{\hbox{$^{ 36 }$Cl}}
\def\HI{\hbox{H$^ \mathrm {o }$}}
\def\HeI{\hbox{He$^ \mathrm {o }$}}
\newcommand{\deeg}{$^\circ$}
\def\kms{\hbox{km s$^\mathrm {-1}$}}
\def\cc{\hbox{cm$^{-3}$}}
\def\cmtwo{\hbox{cm$^{-2}$}}
\def\Rs{\hbox{R$_\mathrm {S}$}}

\setcounter{chapter}{0}
\articletitle[Introduction:  Paleoheliosphere versus PaleoLISM]{Introduction: \\ Paleoheliosphere versus PaleoLISM}
\chaptitlerunninghead{Paleoheliosphere versus PaleoLISM}
\author{Priscilla C. Frisch}
\affil{University of Chicago}
\email{frisch@oddjob.uchicago.edu}

\begin{abstract}
Speculations that encounters with interstellar clouds modify the
terrestrial climate have appeared in the scientific literature for
over 85 years.  The articles in this volume seek to give substance to
these speculations by examining the exact mechanisms that link the
pressure and composition of the interstellar medium surrounding the
Sun to the physical properties of the inner heliosphere at the Earth.
\end{abstract}


\section{The Underlying Query}

If the solar galactic environment is to have a discernible effect on
events on the surface of the Earth, it must be through a subtle and
indirect influence on the terrestrial climate.  The scientific and
philosophical literature of the 18th, 19th and 20th centuries all
include discussions of possible cosmic influences on the terrestrial
climate, including the effect of cometary impacts on Earth
(\nolinebreak \cite{Halley:1694}), and the diminished solar radiation
from sunspots, which Herschel attributed to ``holes'' in the luminous
fluid on the surface of the Sun\footnote{In this same paper Herschel
commented that ``Whatever fanciful poets might say, in making the sun
the abode of blessed spirits, or angry moralists devise, in pointing
it out as a fit place for the punishment of the wicked, it does not
appear that they had any other foundation for their assertions than
mere opinion and vague surmise; but now I think myself authorized,
\emph{upon astronomical principles,} to propose the sun as an
inhabitable world, and am persuaded that the foregoing observations,
with the conclusions I have drawn from them, are fully sufficient to
answer every objection that may be made against it.  ''  These
comments show that valuable data are not always interpreted
correctly.} (\nolinebreak \cite{Herschel:1795}). The discovery of interstellar
material in the 20th century led to speculations that encounters with
dense clouds initiated the ice ages (\nolinebreak
\cite{Shapley:1921}), and many papers appeared that explored the
implications of such encounters, including the influence of
interstellar material (ISM) on the interplanetary medium and planetary
atmospheres
(e.g. \cite{Fahr:1968,BegelmanRees:1976,McKayThomas:1978,Thomas:1978,McCrea:1975,TalbotNewman:1977,Willis:1978,ButlerNewmanTalbot:1978}).
The ISM-modulated heliosphere was also believed to affect climate
stability and astrospheres
(e. g. \cite{Frisch:1993a,Frisch:1997,ZankFrisch:1999}).  Recent
advances in our understanding of the solar wind and heliosphere
(e. g. \cite{WangRichardson:2005,Fahr:2004}) justify a new look at
this age-old issue.  This book addresses the underlying question:
\begin{verse}
\emph{How does the heliospheric interaction} \emph{with the interstellar medium
affect the heliosphere, interplanetary medium, and Earth?}
\end{verse}

\begin{figure}
\centering
\parbox[t]{3.10in}{\centering
  \centerline{
\psfig{figure=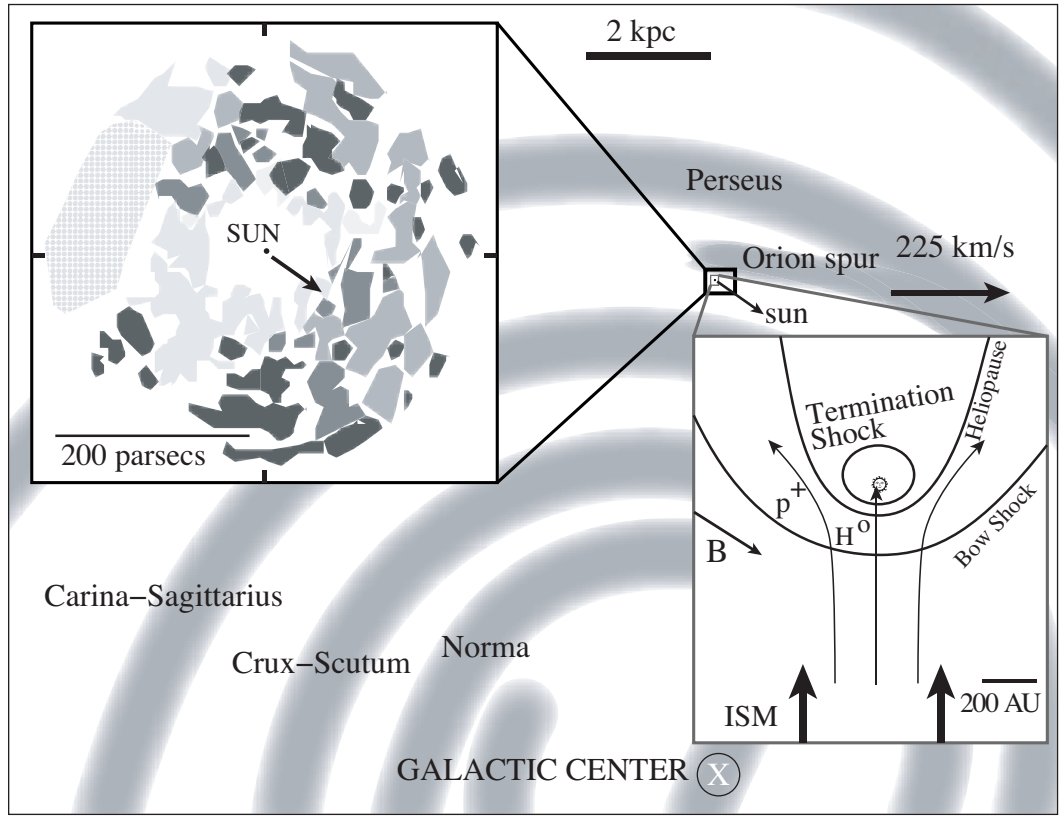,width=5.5in}
}}
\caption[The Solar Journey.]{\label{fig:1}
The solar location and vector motion are identified for the kiloparsec
scale sizes of the Milky Way Galaxy (large image), and for the
$\sim$500 parsec scale size of the Local Bubble (medium sized image, inset in
upper left hand corner).  A schematic drawing of the heliosphere
(small image, inset in lower right hand inset, based on
Linde,1998\nocite{Linde:1998}) shows the upwind velocity of the
interstellar wind (``ISM'') as observed in the rest frame of the Sun.
Coincidently, this direction, which determines the heliosphere nose,
is close to the galactic center direction. The orientation of the
plane in the small inset differs from the planes of the large and
medium figures, since the ecliptic plane is tilted by 60$^\circ$ with
respect to the galactic plane.  The Sun is 8 kpc from the center
of the Milky Way Galaxy, 
and the solar neighborhood moves towards the direction
\glong=90\deeg\ at a velocity of 225 \kms.  The spiral arm positions
are drawn from Vallee (2005), except for the Orion
spur. \nocite{ValleeSA:2005} The Local Bubble configuration is based
on measurements of starlight reddening by interstellar dust (Chapter
6).  The lowest level of shading corresponds to color excess values
\ebv=0.051 mag, or column densities log $N$(H) (\cmtwo) =20.40 dex.
The dotted region shows the widespread ionized gas associated with the
Gum Nebula.  The heliosphere cartoon shows interstellar protons
deflected in the plasma flow in the outer heliosheath regions,
compared to the interstellar neutrals that penetrate the heliopause.}
\end{figure}

The heliosphere is the cavity in the interstellar medium created by
the dynamic ram pressure of the radially expanding solar wind, a halo
of plasma around the Sun and planets, dancing like a candle in the
wind and regulating the flux of cosmic rays and interstellar material
at the Earth.  Neutral interstellar gas and large interstellar dust
grains penetrate the heliosphere, but the solar wind acts as a buffer
between the Earth and most other interstellar material and low energy
galactic cosmic rays (GCR).  Together the solar wind and interstellar
medium determine the properties of the heliosphere.  In the present
epoch the densities of the solar wind and interstellar neutrals are
approximately equal outside of the Jupiter orbit.  Solar activity
levels drive the heliosphere from within, and the physical properties
of the surrounding interstellar cloud constrain the heliosphere from
without, so that the boundary conditions of the heliosphere are set by
interstellar material.  Figure \ref{fig:1} shows the Sun and
heliosphere in the setting of the Milky Way Galaxy.

The answer to the question posed above lies in an interdisciplinary
study of the coupling between the interstellar medium and the solar
wind, and the effects that ISM variations have on the 1 AU environment
of the Earth through this coupling.  The articles in this book explore
different viewpoints, including \emph{gedanken} experiments, as well
as data-rich summaries of variations in the solar environment and
paleoclimate data on cosmic ray flux variations at Earth.  

The book begins with the development of theoretical models of the
heliosphere that demonstrate the sensitivity of the heliosphere to the
variations in boundary conditions caused by the passage of the Sun
through interstellar clouds (\cite{Zanketal:2006jos,PogorelovZank:2006jos}).  A series of \emph{gedanken} experiments
then yield the response of planetary magnetospheres to encounters with
denser ISM (\cite{Parker:2006jos}).  Variations in the galactic environment of the Sun, caused
by the motions of the Sun and clouds through the Galaxy, are shown to
occur for both long and short timescales (\cite{Shaviv:2006jos,FrischSlavin:2006jos}).

The heliosphere acts as a buffer between the Earth and interstellar
medium, so that dust and particle populations inside of the
heliosphere, which have an interstellar origin, vary as the Sun
traverses interstellar clouds.  These buffering mechanisms determine
the interplanetary medium\footnote{The buffering processes convert
interstellar neutrals into low energy ions, which are convected
outwards with the solar wind and accelerated to low cosmic ray
energies that have an anomalous composition, including abundant
elements with FIP$>$13.6 eV. The high energy galactic cosmic ray
population incident on the heliosphere is also modulated.}.  The
properties of these buffering interactions are evaluated for
heliosphere models that have been developed using boundary conditions
appropriate for when the Sun traverses different types of interstellar
clouds (\cite{Landgraf:2006jos,Moebiusetal:2006jos,FlorinskiZank:2006jos,Fahretal:2006jos}).

The consequences of Sun-cloud encounters are then discussed in terms
of the accretion of ISM onto the terrestrial atmosphere for dense
cloud encounters, and the possibly extreme variations expected for
cosmic ray modulation when interstellar densities vary
substantially (\cite{Fahretal:2006jos,YeghikyanFahr:2006jos}). Radioisotope records on Earth extending backwards in
time for over $\sim$0.5 Myrs, together with paleoclimate data, suggest
that cosmic ray fluxes are related to climate.  The galactic
environment of the Sun must have left an imprint on the geological
record through variations in the concentrations of radioactive
isotopes (\cite{KirkbyCarslaw:2006jos}).

The selection of topics in this book is based partly on scientific
areas that have already been discussed in the literature.  The authors
who were invited to contribute chapters have previously studied the
heliosphere or terrestrial response to variable ISM conditions or cosmic rays.

Figure \ref{fig:1} shows the heliosphere in our setting of the Milky
Way Galaxy.  A postscript at the end of this chapter lists basic useful
information.  I introduce the term ``paleoheliosphere'' to represent the
heliosphere in the past, when the boundary conditions set by the local 
interstellar material (LISM) may have differed substantially from the boundary
conditions for the present-day heliosphere.  The ``paleolism'' is the
local ISM that once surrounded the heliosphere.

\section[Addressing the Query: The Heliosphere for Different Interstellar Environments]{Addressing the Query:  The Heliosphere and Particle Populations
for Different Interstellar Environments}

The solar wind drives the heliosphere from the inside, with the
properties of the solar wind varying with ecliptic latitude and the
phase of the 11-year solar activity cycle.  The global heliosphere is
the volume of space occupied by the supersonic and subsonic solar
wind.  Interstellar material forms the boundary conditions of the
heliosphere, and the windward side of the heliosphere, or the ``upwind
direction'', is defined by the interstellar velocity vector 
with respect to the Sun.  The leeward side of the heliosphere
is the ``downwind direction''.  Figure \ref{fig:1} shows a cartoon of
the present-day heliosphere, with labels for the major landmarks such
as the termination shock, heliopause, and bow shock.

In the present-day heliosphere, the transition from solar wind to
interstellar plasma occurs at a contact discontinuity known as the
``heliopause'', which is formed where the total solar wind and
interstellar pressures equilibrate (\nolinebreak \cite{Holzer:1989}).
For a non-zero interstellar cloud velocity in the solar rest frame,
the solar wind turns around at the heliopause and flows around the
flanks of the heliosphere and into the downwind heliotail.  Before
reaching the heliopause, the supersonic solar wind slows to subsonic
velocities at the ``termination shock'', where kinetic energy is
converted to thermal energy.

The subsonic solar wind region between the termination shock and
heliopause is called the inner ``heliosheath''.  The outer heliosheath lies
just beyond the heliopause, where the pristine ISM is distorted by the
ram pressure of the heliosphere.  A bow shock, where the interstellar
gas becomes subsonic, is expected to form ahead of the present-day
heliosphere in the observed upwind direction of the ISM flow through
the solar system.  

Large interstellar dust grains and interstellar atoms that remain
neutral inside of the orbit of Earth, such as He, are gravitationally
focused in the downwind direction.  This ``focusing cone'' is traversed 
by the Earth every year in early December, and extends many AU from the Sun
in the leeward direction
(e.g. \cite{Landgraf:2000,Moebiusetal:2004,Frisch:2000amsci}).  The
heliotail itself extends $>10^3$ AU from the Sun in the downwind
direction, forming a cosmic wake for the solar system.  

Of significance when considering the interaction of the heliosphere
with an interstellar cloud is that neutral particles enter the
heliosphere relatively unimpeded, after which they are ionized and
convected outwards with the solar wind.  Ions and small charged dust
grains are magnetically deflected in the heliosheath around the flanks
of the heliosphere (see Figure \ref{fig:1}).

Space and astronomical data now confirm the basic milestones of the
outer heliosphere.  Voyager 1 crossed the termination shock at 94 AU
on 16 December, 2004 (UT), and observed the signature of the
termination shock on low-energy particle populations, the solar wind
magnetic field, low-energy electrons and protons, and Langmuir radio
emission (\nolinebreak
\cite{Stoneetal:2005,Burlagaetal:2005,Gurnett:2005,Deckeretal:2005}).
The present-day termination shock appears to be weak, with a solar
wind velocity jump ratio (the ratio of upstream to downstream values)
of $\sim$2.6 and a magnetic field compression ratio of $\sim$3.  The
magnetic wall that is predicted for the heliosphere (\nolinebreak
\cite{Linde:1998,Ratkiewicz:1998}, Chapter 3 by Pogorelov and Zank)
appears to have been detected through observations of magnetically
aligned dust grains (\nolinebreak
\cite{Frisch:2005L}), and the offset between upwind directions of
interstellar \HI\ and \HeI\ (\nolinebreak \cite{Lallementetal:2005}).
The compressed and heated \HI\ in the hydrogen wall region of the
outer heliosheath has now been detected around a number of stars
(\nolinebreak \cite{Woodetal:2005}).

The present-day solar wind is the baseline for evaluating the
heliosphere response to ISM variations in the following articles, so a
short review of the solar wind is first presented.  The remaining part
of $\S$1.2 introduces the topics in the following articles in terms of
the underlying query of the book.

\subsection{The Present Day Solar Wind }

The solar wind originates in the million degree solar corona that
expands radially outwards, with a density $\sim 1/R^2_\mathrm{ S }$
where $R_\mathrm{ S }$ is the distance to the Sun, and contains both
features that corotate with the Sun, and transient structures (\nolinebreak
e.~g.~\cite{Gosling:1996}).  The properties of the solar wind vary
with the phase of the solar magnetic activity cycle and with ecliptic
latitude.  The best historical indicator of solar magnetic activity
levels is the number of sunspots, first detected by Galileo in 1610,
which are magnetic storms in the convective zone of the Sun.  Sunspot
numbers indicate that the magnetic activity levels fluctuate with a
$\sim$11 year cycle, or the ``solar cycle'', and solar maximum/minimum
corresponds to the maximum/minimum of sunspot numbers.  The magnetic
polarity of the Sun varies with a $\sim$22 year cycle.  During solar
maximum, a low-speed wind, with velocity $\sim$300--600 \kms\ and
density $\sim 6-10$ particles \cc\ at 1 AU, extends over most of the
solar disk.  Open magnetic field lines\footnote{Open magnetic field lines
are formed in coronal holes that reconnect in the outer heliosphere
and contain low density and very high speed, $\sim$700 \kms, solar wind.} 
are limited to solar pole regions.  A neutral current sheet $\sim$0.4 AU 
thick forms between the solar wind containing negative magnetic polarity 
fields and the solar wind that contains positive magnetic polarity fields.
The neutral current sheet reaches its largest 
inclination ($\ge 70^\circ$) during solar maximum.
During the conditions of solar minimum, a high speed
wind with velocity $\sim$600--800 \kms\ and density $\sim$5 \cc\ is
accelerated in the open magnetic flux lines in coronal holes.  During
mininum, the
high speed wind and open field lines extend from the polar regions
down to latitudes of $\leq 40$\deeg
(\nolinebreak \cite{Smithetal:2003,Richardson:1995}).  The higher solar wind
momentum flux associated with solar minimum conditions produces an
upwind termination shock that is $\sim$5--40 AU more distant in the
upwind direction than during solar maximum conditions
(e.g. \cite{SchererFahr:2003,ZankMueller:2003,Whang:2004}).

The alignment and strength of the solar magnetic multipoles depend on the phase of the solar cycle
(\cite{Bravoetal:1998}).  During solar minimum conditions, the magnetic field 
is dominated by the dipole and hexapole moments, and the dipole moment
is generally aligned with the solar rotation axis.  
Sunspots migrate from high to low heliographic latitudes.
The magnetic poles follow the coronal holes to the solar equator as solar
activity increases.  During the solar maximum period, the galactic cosmic 
rays undergo their maximum modulation, the dipole component of the
magnetic field is minimized, the quadrapole moment dominates, and the polarity of the solar magnetic field 
reverses (\cite{LockwoodWebber:2005}, Figure
\ref{fig:2}).  Over historic times, the cosmic ray modulation by 
the heliosphere correlates better with the open magnetic flux line 
coverage than with sunspot numbers (\cite{McCrackenetal:2004}).

Variable cosmic ray modulation produced by a variable heliosphere may
be a primary factor in both solar and ISM forcing of the terrestrial
climate.  The heliosphere modulation of cosmic rays is well
established.  John Simpson, to whom this book is dedicated, initiated
a program 5 solar cycles ago in 1951 to monitor cosmic ray fluxes on
Earth using high-altitude neutron detectors (\cite{Simpson:2001}).
The results show a pronounced anticorrelation between cosmic ray flux
levels and solar sunspot numbers, which trace the 11-year Schwabe
magnetic activity cycle, and which also show that the polarity of the solar
magnetic field affects cosmic ray modulation (see Figure \ref{fig:2}).
The articles in this book show convincingly that the ISM also modulates the 
heliosphere, and the effect of the solar wind on the heliosphere
must be differentiated from the influence of interstellar matter.

Variations in solar activity levels are also seen over $\sim 100-200$
year timescales, such as the absence of sunspots during the Maunder
Minimum in the 17th century.  Modern climate records show that the
Maunder Minimum corresponded to extremely cold weather, and 
radioisotope records show that the flux of cosmic rays
was unusually high at this time (see Kirkby and Carslaw, Chapter 12).
Similar effects will occur from the modulation of galactic cosmic rays
by the passage of the Sun through an interstellar cloud.

These temporal and latitudinal variations in the solar wind momentum
flux produce an asymmetric heliosphere, which varies with time.  Any possible
historical signature of the ISM on the heliosphere must first be
distinguished from variations driven by the solar wind itself.

\subsection{Present Day Heliosphere and Sensitivity to ISM}

The ISM forms the boundary conditions of the heliosphere, so that
encounters with interstellar clouds will affect the global
heliosphere, the interplanetary medium, and the inner heliosphere
region where the Earth is located.  Today an interstellar wind passes
through the solar system at --26.3 \kms\ (\cite{Witte:2004}).  An
entering parcel of ISM takes about 20 years to reach the inner
heliosphere, so that ISM near the Earth is constantly replenished with
new inflowing material.  This warm gas is low density and partially
ionized, with temperature $T \sim$6,300 K, and densities of neutral
and ionized matter of \nHI$\sim 0.2$ \cc, and
\nHII$\sim 0.1$ \cc.

An elementary perspective of the response of the heliosphere to
interstellar pressures is given by an analytical expression for the
heliopause distance based on the locus of positions where the solar
wind ram pressure, $P_\mathrm{SW}$, and the total interstellar
pressure equilibrate (\cite{Holzer:1989}).  The solar wind density
$\rho$ falls off as $\sim 1/R^2$, where $R$ is the distance to the
Sun, while the velocity $v$ is relatively constant.  At 1 AU the solar
wind ram pressure is $P_\mathrm{SW,1AU} \sim \rho~v^2 $ so the
heliosphere distance, $R_\mathrm{HP} $, is given by:

\begin{eqnarray*}
P_\mathrm{SW,1AU}/R_\mathrm{HP}^2  \sim  P_\mathrm{B} +P_\mathrm{Ions, thermal} +
P_\mathrm{Ions, ram} + P_\mathrm{Dust} + P_\mathrm{CR} 
\end{eqnarray*}
The interstellar pressure terms include the magnetic pressure
$P_\mathrm{B}$, the thermal, $P_\mathrm{Ions, thermal}$, and the ram,
$P_\mathrm{Ions, ram}$, pressures of the charged gas, and the
pressures of dust grains, $P_\mathrm{Dust}$, and cosmic rays,
$P_\mathrm{CR}$, which are excluded by heliosphere magnetic fields and
plasma.  Some interstellar neutrals convert to ions through charge
exchange with compressed interstellar proton gas in heliosheath
regions, adding to the confining pressure.  An important response
characteristic is that, for many clouds, the encounter will be
ram-pressure dominated, where $P_\mathrm{ram} \sim m v^2$ for
interstellar cloud mass density $m$ and relative Sun-cloud velocity
$v$, so that variations in the cloud velocity perturb the heliosphere
even if the thermal pressures remain constant.

The multifluid, magnetohydrodynamic (MHD), hydrodynamic and hybrid
approaches used in the following chapters provide much more
substantial models for the heliosphere, and include the coupling
between neutrals and plasma, and field-particle interactions.  These
sophisticated models predict variations in the global heliosphere in
the face of changing interstellar boundary conditions, and for a range
of different cloud types.  Although impossible to model a solar
encounter with every type of interstellar cloud, the following
articles include discussions of many of the extremes of the
interstellar parameter space, including low density gas with a range
of velocities, very tenuous plasma, high velocity clouds, dense ISM,
and magnetized material for a range of field orientations and
strengths.  The discussions in these chapters extrapolate from our
best theoretical understanding of the heliosphere boundary conditions
today to values that differ, in some cases dramatically, from the
boundary conditions that prevailed at the beginning of the third
millennium in the Gregorian calendar.

The Sun has been, and will be, subjected to many different
physical environments over its lifetime. Theoretical heliosphere
models yield the properties
of the solar wind-ISM interaction for these different environments,
which in turn determine the nature and properties of interstellar
populations inside of the heliosphere for a range of galactic environments.
These models form the foundation for understanding the significance
of our galactic environment for the Earth.

The interstellar parameter space is explored
by Zank et al. (Chapter 2), where 28 sets of
boundary conditions are evaluated with computationally efficient multifluid
models.  Moebius et al. (Chapter 8), Fahr et al. (Chapter 9),
Florinski and Zank (Chapter 10), and Yeghikyan and Fahr (Chapter 11) also develop heliosphere models
for a range of interstellar conditions.  Together these models 
evaluate the heliosphere response to interstellar density, temperature, 
and velocity variations of factors of $\sim 10^{9}$, $\sim 10^{5}$, and $\sim 10^{2}$,
respectively.

The interstellar magnetic field introduces an asymmetric pressure
on the heliosphere, affecting the heliosphere current sheet
and cosmic ray modulation.
Pogorelov and Zank (Chapter 3) use MHD models to probe the heliosphere
response to the interstellar magnetic field,
including charge exchange between
the neutrals and solar wind.  The resulting asymmetry provides a 
test of the magnetic field direction, and shows strong
differences between cases where the interstellar flow is parallel,
instead of perpendicular, to the interstellar magnetic field direction.
Since the random component of the interstellar magnetic field
is stronger, on the average, than the ordered component, particularly
in spiral arm regions where active star formation occurs,
a range of interstellar magnetic field strengths and
orientations are expected over the solar lifetime
(Shaviv, Chapter 5, and Frisch and Slavin, Chapter 6).

\subsection{Planetary Magnetospheres}

The Earth's magnetosphere acts as a buffer between the solar wind and
atmosphere, and as such is an ingredient in understanding the effect
of our galactic environment on the Earth.  The decreasing solar wind
density in the outer heliosphere results in an interplanetary medium
around outer planets that is more sensitive to ISM variations than for
inner planets, with implications for the magnetospheres of Jupiter,
Neptune, and Uranus.  Most topics in this book are already considered
in the scientific literature, but questions about magnetosphere
variations from an ISM-modulated heliosphere have received scant
attention.  In a quintessential \emph{gedanken} experiment, Parker
explores the interaction between magnetospheres and the solar wind for
variations in the interstellar density, and for inner versus outer
planets (Chapter 4).

\subsection{Short and Long Term Variations in the Galactic Environment}

There is every reason to expect that the galactic environment of the
Sun varies over geological timescales.  The Sun moves through space at
a velocity of 13--20 \kms, and interstellar clouds have velocities
ranging up to hundreds of \kms.  The Arecibo Millennium survey showed
that $\sim$25\% of the mass contained in interstellar \HI, including
both warm and cold ISM, is in clouds traveling with velocities $\ge$10
\kms\ through the local standard of rest (\nolinebreak \cite{HTI}).
Thus Sun-cloud encounters with relative velocities exceeding 25 \kms\
are quite likely, and for a typical cloud length of $\sim$1 pc the
cloud transit time would be $\sim$40,000 years.  The many types of ISM
traversed by the Sun during the past several million years have
affected the heliosphere, the inner solar system, and the flux of
anomalous and galactic cosmic rays at Earth
(\cite{FrischYork:1986,Frisch:1997,Frisch:1998}).

For the past $\sim$3 Myrs the Sun has been in a nearly empty region of
space, the ``Local Bubble'', with very low densities of $<$10$^{-26}$
gr \cc.  Within the past 44,000--140,000 years the Sun entered a
flow of tenuous, partly neutral ISM, nick-named the ``Local Fluff'', with density
$\sim$60 times higher (Chapter 6).  This transition was accompanied by
the appearance of interstellar dust and neutrals in the heliosphere,
along with the pickup ion and anomalous cosmic ray populations.
Galactic cosmic ray modulation was affected, providing a possible link
between our galactic environment and climate.  Intriguingly, the
averaged cosmic ray flux at Earth, as traced by \Beten\ records, was
lower in the past $\sim$135 kyrs than for earlier times (Chapter 12).
Was the decrease in the galactic cosmic ray flux $\sim$135 kyrs ago
caused by an increase in modulation as the Sun entered the Local
Fluff?

The galactic environment of the Sun also varies quite dramatically over
long time scales, as discussed by Shaviv (Chapter 5).  Over its 4.5 billion year lifetime,
the Sun traverses spiral arm and interarm regions, with atomic
densities varying from less than $10^{-26.1}$ g \cc\ to over
$10^{-20.1}$ g \cc, and temperatures ranging over 7 orders of
magnitude, 10--10$^7$ K.  The Sun is now in low density space between
the Perseus and Sagittarius spiral arms, and on the inner edge of what
is known as the Orion Spur on the Local Arm.  The Local Arm is not
shown in Figure \ref{fig:1}, as is consistent with the usual Galaxy
depictions.  The Local Arm does not appear to be a grand design spiral shock (Bochkarev,
1984\nocite{Bochkarev:1984}).  The Sun has a systematic motion of
13--20 km s$^{-1}$ with respect to the nearest stars, corresponding to
$\sim$3--4 AU per year.  The Local Interstellar Cloud (LIC) now
surrounding the Sun traverses the heliosphere at $\sim$5.5 AU per
year.  The Sun oscillates vertically through the galactic plane once
every $\sim$34 Myrs, and orbits the center of the Milky Way Galaxy 
once per $\sim$220 Myrs.  

Shaviv evaluates variations in the galactic environment of the Sun
over long timescales.  This bold discussion compares
various geologic records of cosmic ray flux variations, based on
radioisotope data that sample timescales of $\sim 10^8$ years, with
models of the Milky Way Galaxy spiral arm pattern to reconstruct the
timing of the Sun's passage through spiral arms.  The chapter concludes that star
formation in spiral arms leaves a signature on the radioisotope
records of the solar system.

Frisch and Slavin (Chapter 6) reconstruct short-term variations of the
galactic environment of the Sun using observations of interstellar
matter towards nearby stars and inside of the solar system.  Radiative
transfer models of the LIC show that ionization varies across
this low density cloud, so that the heliosphere boundary conditions vary
from radiative transfer considerations alone as the Sun traverses the LIC.
Cloud transitions are predicted during the past $\sim$3 Myrs, including the
departure of the Sun from the Local Bubble interior 44,000--140,000
years ago, and entry into the surrounding cloud 1000--40,000 years ago.

\subsection{Interstellar Dust}

The particle populations formed by the interactions between the solar
wind and interstellar dust, gas, and cosmic rays are emissaries
between the cosmos and inner heliosphere, varying as the Sun moves
through clouds.

About $\sim$1\% of the mass of the cloud surrounding the Sun is
contained in interstellar dust grains.  The largest of these charged
grains, mass $> 10^{-13} $ g, have large magnetic Larmor radii of
$>$500 AU at the heliopause for an interstellar field of $\sim$1.5
$\mu$G, and flow into the solar system.  The Earth passes through the
gravitational focusing cone formed by these grains early each
December.  The smallest charged grains, mass$< 10^{-14.5} $ g and
radii$< 0.01 ~\mu$m, have Larmor radii of $\sim$20 AU, depending on
the magnetic field strength and radiation field, and are deflected
around the heliosheath (\cite{Frischetal:1999}).  Interstellar
dust grains are measured in
the inner heliosphere within $\sim$5 AU of the Sun, and over the solar
poles, by satellites such as Ulysses, Galileo and Cassini.  Landgraf
(Chapter 7) reviews the properties of the interaction between
interstellar dust and the solar wind, and speculates on the changes
that might be expected from an encounter with a dense interstellar
cloud.

Should it some day be possible to
compare the ratio of large to small interstellar dust grains on the
surfaces of the inner versus outer planets, it would become possible to
disentangle cloud encounters from solar activity effects.

At the very large end of the dust population mass spectrum we find
interstellar micrometeorites, with masses $\sim$3 $\times$ 10$^{-7}$
g, open orbits, and inflow velocities greater than the 42 \kms\ escape
velocity from the solar system at 1 AU.  These interstellar objects,
detected by radar as they impact the atmosphere, evidently originate
in circumstellar disks such as that around $\beta$ Pictoris, and in the
interior of the Local Bubble (\cite{Baggaley:2000,Meiseletal:2002}).
These objects do not collisionally couple to the interstellar gas
(\cite{GruenLandgraf:2000}), and should not vary with the type of ISM
surrounding the Sun.

\subsection{Particle Populations in the Inner and Outer Heliosphere}

Presently, low energy interstellar neutrals, high energy galactic
cosmic rays, and interstellar dust all enter the heliosphere.
The characteristics of each of these populations and their secondary
products are modified as the Sun transits the ISM, or the cloud
ionization changes.
The first ionization potential (FIP) of \HI\ is 13.6 eV.  Neutral
interstellar atoms with FIP$<$13.6 eV are ionized in nearly all
interstellar clouds because the main source of interstellar opacity is
\HI.  Interstellar ions are deflected around the heliosheath, so
the result is that only interstellar atoms with FIP$>$13.6 eV
enter the heliosphere where they are then destroyed, primarily by charge
exchange with solar wind ions.  

The density of interstellar neutrals in the inner heliosphere 
depends on the density and ionization of the surrounding cloud, the
ionization (or ``filtration'') of those neutrals by the heliosheath,
and the subsequent interactions with the solar wind inside of the heliosphere.
Secondary products produced by solar wind interactions
with interstellar neutrals inside of the heliosphere 
include pickup ions 
\footnote{The pickup ions are interstellar neutrals formed by charge
exchange with the solar wind.  Energetic neutral atoms are formed by
energetic ions that capture an electron from a low energy neutral by
charge exchange.  The gravitational focusing cone contains heavy
elements (mainly He) that are predominantly ionized inside of 1 AU and
therefore gravitationally focused downwind of the Sun (Chapter 8).
Large interstellar dust grains are also gravitationally focused (Chapter
7).  The anomalous cosmic ray population is formed from pickup ions
accelerated to low cosmic ray energies, $<$\nolinebreak 1 GeV, in the
solar wind and at the termination shock, and then subjected to the
same modulation and propagation processes as galactic cosmic rays
(\cite{Jokipii:2004}).}, energetic neutral atoms, the gravitational
focusing cone formed by helium (also seen in dust), and the anomalous cosmic ray
population with energies $<$\nolinebreak 1 GeV.  Interstellar neutrals
inside of the heliosphere, and the heliosphere itself, form a coupled
system that together respond to variations in the heliosphere boundary
conditions.

Moebius et al. (Chapter 8) model the heliosphere for several
different conditions, and then probe the response of the inner
heliosphere to the density of interstellar neutrals flowing
into this ISM-modified heliosphere.  At 1 AU, the neutral densities, particle
populations derived from interstellar neutrals,  and 
characteristics of the helium focusing cone all respond to
variations in the interstellar boundary conditions.
For some cases, increased neutral fluxes fall on the atmosphere
of Earth (also see Yeghikyan and Fahr, Chapter 11).

The velocity structure of the ISM appears to vary on subparsec scale
lengths (Frisch and Slavin, Chapter 6), and these variations may in 
some cases result in significant modifications of the inner heliosphere, 
particularly the gravitational focusing cone, when all other interstellar parameters 
such as thermal pressure are invariant (Zank et al., Chapter 2, Moebius 
et al., Chapter 8).

The most readily available diagnostics of the paleoheliosphere are
radioisotopes, formed by cosmic ray spallation on the atmosphere,
interplanetary and interstellar dust, and meteorites.  Thus,
the evaluation of cosmic ray modulation for various types of
interstellar cloud boundary conditions is a key part of understanding
the paleoclimate records that might trace the solar journey
through the Milky Way Galaxy. Fahr et
al. (Chapter 9) and Florinski and Zank (Chapter 10) use our
understanding of galactic cosmic ray modulation in the modern-day
heliosphere as a basis for making detailed calculations of the
response of the paleoheliosphere, or the heliosphere as it once was,
to the paleolism, or the local interstellar medium that once
surrounded the Sun.  The predictions of these calculations are quite
intriguing.  Both the termination shock compression ratio and the
solar wind turbulence spectrum may vary dramatically with different
environments, as mass-loading by pickup ions and the heliosphere
properties vary.  The problem of galactic cosmic ray modulation in an
ISM-forced heliosphere is extremely important to understanding the
paleoheliosphere signature in the terrestrial isotope record.

Today, galactic cosmic rays (GCR) with energies $\ge 0.25$ GeV
penetrate the solar system, and anomalous cosmic rays 
(energies $<1$ GeV) are formed from accelerated pickup ions.  
The cosmic ray flux at Earth is sampled by geological radioisotope records,
as reviewed Kirkby and Carslaw (Chapter 12, also see Florinski and Zank, Chapter 10).  
Astronomical data indicates that the Sun has emerged from a region of space with
virtually no neutral ISM within the past $\sim$0.4--1.5 10$^5$ years,
and entered the Local Fluff (Chapter 6).  The GCR modulation
discontinuity that accompanied this transition may be in the
geologic record, which show lower cosmic ray fluxes at Earth,
on the average, for the past 135 kyrs years
than the 135 kyrs before that (\cite{Christl:2004}).

\subsection{Atmosphere Accretion from Dense Cloud Encounters}

Harlow Shapley (1921) suggested \nocite{Shapley:1921} that an
encounter between the Sun and giant dust clouds in Orion may have
perturbed the terrestrial climate and caused ice ages.  The discovery
of interstellar \HI\ and \HeI\ inside the heliosphere was soon
followed by studies of the ISM influence on the atmosphere for dense
cloud conditions
(\cite{Fahr:1968,BegelmanRees:1976,McKayThomas:1978,Thomas:1978,McCrea:1975,TalbotNewman:1977,Willis:1978,ButlerNewmanTalbot:1978}).
Yeghikyan and Fahr (Chapter 11), evaluate the density of ISM at the
Earth based on models describing the heliosphere inside of an dense cloud, 
and the interactions between the solar wind and ISM for these dense
cloud conditions (also see Chapter 9, by Fahr et al.).  These models 
then yield the concentration of interstellar hydrogen at the Earth, 
and the flow of water downward towards the Earth's
surface, as a function of the dense cloud density.
Significant atmosphere modifications are predicted in some cases.  
Enhanced neutral populations at 1 AU for a somewhat lower interstellar cloud density regime 
are discussed in Chapter 8, by Moebius et al.

\subsection{Possible Effects of Cosmic Rays}

Both solar activity cycles (Figure \ref{fig:2}) and ISM variations
modulate the cosmic ray flux in the heliosphere, and Kirkby and
Carslaw (Chapter 12) compare galactic cosmic ray records with
paleoclimate archives.  They examine sources of climate forcing such
as solar irradiance and cosmic ray fluxes, and conclude that arguments
in favor of cosmic ray climate forcing are strong although the
mechanism is uncertain.  This relation between cosmic ray flux levels
and the climate is shown by radioisotope records and climate archives,
such as ice cores, stalagmites, and ice-rafted debris, and for modern
times, by historical records. Paleoclimate archives include
terrestrial records of cosmic ray spallation in the atmosphere, as
traced by radioisotopes with short half-lives (\tauhalf), e.~g.~
\Cfourteen\ (\tauhalf=5,730 yrs) and \Beten\ (\tauhalf=1.6 Myrs).
Possible mechanisms linking the cosmic ray flux at 1 AU and the
climate include cloud nucleation by cosmic rays, and the global
electrical circuit (see Chapter 12 and \cite{RobleHaysII:1979}).  The discussion in
Chapter 12 provides persuasive evidence linking the
surface temperature to cosmic ray fluxes at Earth.  The
anticorrelation between sunspot number and cosmic ray fluxes in Figure
\ref{fig:2} shows the heliosphere role in cosmic ray modulation; 
this mechanism must have also been a prominent mechanism for relating
the ISM-modulated heliosphere with the climate.  Fortunately this
hypothesis is also verifiable by comparing paleoclimate data with
astronomical data on the timing of cloud transitions.

The radioisotope records also indicate that cosmic ray fluctuations
have occurred over longer timescales of many $10 ^8$ years.
Shaviv compares the \Clthirtysix\ (\tauhalf $\sim$0.3 Myrs) and
\Kforty\ (\tauhalf $\sim$1.3 Gyr) cosmic ray exposure records in iron
meteorites (Chapter 5), but in this case to obtain cosmic ray flux
increases due to the Sun's location in spiral arms where active star
formation occurs.

A number of studies, none convincing, have invoked the geological
\Beten\ record, as a proxy for cosmic ray fluxes at Earth, to infer
historical encounters with interstellar clouds.  As a way of dating
the Loop I supernova remnant, it was suggested that the relative
constancy of \Beten\ in sea sediments precluded a strong nearby X-ray
source within the past $\sim$2 Myrs (\cite{Frisch:1981}).  Sonett
(1992) \nocite{Sonett:1992} suggested that peaks in \Beten\ layers
35,000 and 65,000 years ago resulted from a compressed heliosphere
caused by the passage of a high-velocity interstellar shock.  This
extreme heliosphere compression expected for a rapidly moving cloud is
supported by heliosphere models (Chapter 2).  Structure in the \Beten\
peaks has also been related to spatial structure in the local ISM 
(\cite{Frisch:1997}), and solar wind turbulence caused by mass-loading 
of interstellar neutrals may supply the required mechanism.
Global geomagnetic excursions such as the events
$\sim$32 kyr and $\sim$40 kyr ago also affect the \Beten\ record, and
can not be ignored (\nolinebreak \cite{Christl:2004}).  Indeed, Figure
\ref{fig:2} shows the sensitivity of galactic cosmic ray fluxes on
Earth to geomagnetic latitude.

\section{Closing Comments}

This brief summary of the scientific question motivating this
book does not relay the full significance of the galactic environment
of the Sun to the heliosphere and Earth; the following chapters
provide deeper insights into this question.

Historical and paleoclimate data show a correspondence between high
cosmic ray flux levels and cool temperatures on Earth
(\cite{Parker:1996}).  The disappearance of sunspots for extended
periods of time, such as the Maunder Minimum in the years 1645 to
1715, shows up in terrestrial radioisotope records such as \Beten\ in
ice cores (Chapter 12).  The solar magnetic activity cycle was present
during this period, and cosmic ray modulation by the heliosphere was
still evident (\cite{McCrackenetal:2004}).  The \Beten\ record now
extends to $\sim$10$^5$ years before present, raising the hope that
encounters between the Sun and interstellar clouds can be separated
from solar activity effects, and from the global signature of
geomagnetic pole wandering.

Sunspots have long been controversial as an influence on the
terrestrial climate.  Sir William Herschel carefully observed them, and
postulated that diminished solar radiation at Earth during sunspot
maximum affected the terrestrial climate
(1801). \nocite{Herschel:1801} Prof. Langley (1876)
\nocite{Langley:1876} measured the radiative heat from sunspot umbral
and penumbral regions, and concluded the $<$0.1\% solar radiation
decrease associated with sunspots was inadequate to affect the
climate.  Climate records show that the Maunder Minimum and other
periods of low solar activity levels have been exceptionally cold,
which implicates high cosmic ray fluxes with cold climate conditions.
Solar activity levels have returned to historic highs in the past few
decades (\cite{CaballeroMcCracken:2004}), and the historic
correlations indicate these high levels also yield warm climate
conditions.  Unfortunately, these scientific conclusions also impact
the politically loaded issue of global warming.

The possibility that the cosmos has affected the terrestrial climate
is a longtime source of speculation, with many of the first
discussions focused on explaining the ``Universal Deluge".  In 1694
Edmond Halley presented his thoughts to the Royal Society as to
whether the "casual Shock of a Comet, or other transient Body" might
instantly alter the axis orientation or diurnal rotation of the Earth,
thus disturbing sea levels, or whether the impact of a comet could
explain the presence of "vast Quantities of Earth and high Cliffs upon
Beds of Shells, which once were the Bottom of the Sea"
(\cite{Halley:1694}).  Halley's speculation has resurfaced in the
hypothesis that the impact of a comet led to the extinction of
dinosaurs 65 Myrs ago at the Cretaceous-Tertiary boundary
(\nolinebreak \cite{Alvarez:1982}).  The common sense disclaimer that accompanied
Halley's discussion is timeless: \emph{ ``... the Almighty generally
making us of Natural Means to bring about his Will, I thought it not
amiss to give this Honourable Society an Account of some Thoughts that
occurr'd to me on this Subject; wherein, if I err, I shall find myself
in very good Company.''}

The articles in this volume show firmly that the interaction between
the heliosphere and ISM depends on the detailed boundary conditions
set for the heliosphere by each type of interstellar cloud encountered
by the Sun, and that the galactic environment of the Sun changes over
both geologically short time scales of $< 10^5$ years, and long
time scales of $> 10 ^7$ years.  This interaction, in turn, affects
the flux of gas, dust, and energetic particles in the inner
heliosphere.

\begin{figure}[t]
\centering
\parbox[t]{3.10in}{\centering
  \centerline{
\psfig{figure=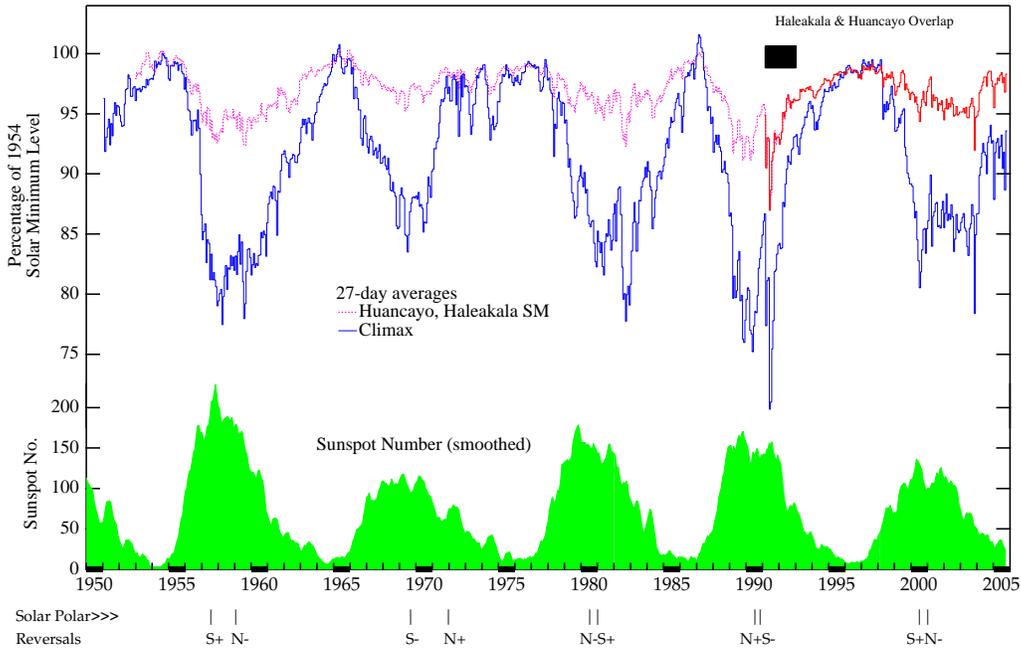,width=5.5in}
}}
\caption[Cosmic Ray Fluxes Versus Sunspot Number.]{\label{fig:2}
Galactic cosmic ray fluxes on Earth versus solar activity levels for 
sunspot cycles 18--23.  
Depicted are 27-day averages of the Climax (blue), and 
Huancayo/Haleakala (pink/red)
neutron monitor rates as a percentage of their respective 1954 solar minimum
levels.  The running averages of the monthly mean sunspot number (green) 
are a proxy for the level of turbulence in the heliosphere as a function of
solar activity.  There is a clear anti-correlation between the neutron monitor
rates and the sunspot number.  The flat-topped versus peaked-top neutron
monitor rates seen at successive 11-year solar minimum periods are a function
of the polarity of the heliospheric magnetic field, noted at the bottom.
The Climax data show solar cycle modulation for $>$3 GeV GCRs, while the 
Huancayo/Haleakala data show solar cycle modulation for $>$13 GeV GCRs
(for additional detail please see \cite{Lopate:2005}).  
The geomagnetic latitudes of Climax and Huancayo/Haleakala are
48$^\circ$ and $\sim 0.5 ^\circ / 20^\circ$.
The poles are given as N and S, for north and south poles.
The terms N+/S- indicate the times when the polarities of the
north/south poles became positive/negative, while N-/S+
indicates they became negative/positive instead.  The
N/S poles do not appear to switch polarities simultaneously.
The author thanks Dr. Clifford Lopate for providing this figure, and
for maintaining a valuable data stream from an experiment begun by 
John Simpson in 1950.  \nocite{Simpson:2001}
}
\end{figure}

The discussions in this book also apply to the study of astrospheres
around cool stars, which are expected to have similar properties as
the heliosphere.  Is the historical astrosphere of a star a factor in
climate stability for planetary systems?  I think so
(\cite{FrischYork:1986}).  If so, then the sample of $\sim$100
detected extrasolar planetary systems can be narrowed to those that
are the most likely to harbor technological civilizations by
evaluating the astrosphere characteristics suitable to the space
trajectory of each star (\nolinebreak \cite{Frisch:1993a}).
Astrospheres have now been detected towards $\sim$60\% of the observed cool
stars within 10 pc (\nolinebreak \cite{Woodetal:2005}), and extensive
efforts to detect Earth-sized exoplanets are underway.  Perhaps some
day these questions will be answered.

\vspace*{0.1in}
\emph{Acknowledgments:}  The author thanks Dr. Clifford Lopate,
of the University of New Hampshire, for providing Figure \ref{fig:2},
and Dr. Lopate thanks NSF Grant 03-39527 for supporting the research
displayed in this figure.  The author thanks NASA for supporting her
research, including grants NAG5-13107 and NNG05GD36G.  Additional support
has been provided by grants NAG 5-13558 and NAG5-11999.  This article
will appear in the book ''Solar Journey: The Significance of Our Galactic Environment 
for the Heliosphere and Earth'',
Springer, in press (2006), editor P. C. Frisch.

\begin{table}[h]
\caption[Commonly Used Terms and Acronyms.]{Commonly Used Terms and Acronyms }
\begin{tabular}{ll}
Object & Description \\
\hline
\emph{Interstellar:}& \\
Interstellar Material, ISM & Atoms in the space between stars \\
Local Fluff or CLIC & ISM within $\sim$30 pc, density $10^{-24.3}$ g \cc \\
& CLIC=Cluster of Local Interstellar Clouds \\
Local Interstellar Cloud, LIC &  The cloud feeding ISM into the solar system \\
Local Bubble, LB & Nearby ISM with density $< 10 ^{-26.1}$ g \cc\\
& \\
\emph{Heliosphere:}& \\
Solar Wind, SW & Solar plasma expanding to form heliosphere \\
& Density$\sim$5 ions \cc, velocity $\sim$450 km s$^{-1}$ \\
& at Earth \\
Neutral Current Sheet & Thin neutral region separating SW \\
& with opposite magnetic polarities \\
Heliosphere, HS  & Region of space containing the solar wind \\
Termination Shock, TS  & Shock where solar wind becomes subsonic \\
& TS at $\sim$94 AU on 16 December, 2004 \\
Heliosheath &  Subsonic solar wind, outside TS \\
Heliosphere Bow Shock & Shock where LIC becomes subsonic \\
Focusing Cone & Gravitationally focused ISM dust \\
& and helium gas downwind of the Sun\\
& \\
\multicolumn{2}{l}{\emph{Interstellar Products in the Heliosphere:}} \\
Pickup Ions, PUI & Ions from SW-ISM charge exchange \\
Energetic Neutral Atoms & ENAs, Energetic atoms formed by \\
 & charge exchange with ions \\
Cosmic Rays:  & \\
\hspace{3mm}Anomalous, ACR & Accelerated pickup ions, energy $<$1 GeV \\
\hspace{3mm}Galactic, GCR & From supernova, energy $>$1 GeV at Earth\\
\hline
\end{tabular}
\end{table}

\section*{Postscript:  Definitions}

The nine planets of the solar system (including Pluto as a planet)
extend out to 39 AU, compared to the distance of the solar wind
termination shock in the upwind direction of 94 AU.  
The Earth is 8.3 light minutes from the Sun, versus the $\sim$0.5 light day
distance to the upwind termination shock of the solar wind.  The
ecliptic and galactic planes are tilted with respect to each other by
$\sim$60$^\circ$, and the north ecliptic pole points towards the
galactic coordinates $\ell$=96.4$^\circ$ and $b$=+29.8$^\circ$.  This
tilt allows the separation of large scale ecliptic and large scale
galactic phenomena by geometric considerations.

Acronyms are used throughout this
book, and some of these are listed in Table 1.
For those new to this subject, an astronomical unit, AU, is the
distance between the Earth and Sun.  A parsec, pc, is 206,000 AU, 3.3
light years (ly), or 3.1 $\times$ 10$^{18}$ cm.  For comparison, the
Earth is 8.3 light minutes from the Sun, and the nearest star,
$\alpha$ Cen, is 1.3 pc from the Sun.  The planet Pluto is at 39 AU,
versus the 94 AU termination shock distance.


\begin{chapthebibliography}{}

\bibitem[{Alvarez}, 1982]{Alvarez:1982}
{Alvarez}, L.~W. (1982).
\newblock {Experimental evidence that an Asteroid Impact LED to the Extinction
  of many Species 65 Million Years Ago}.
\newblock {\em NASA STI/Recon Technical Report N}, 83:33813.

\bibitem[{Baggaley}, 2000]{Baggaley:2000}
{Baggaley}, W.~J. (2000).
\newblock {Advanced Meteor Orbit Radar Observations of Interstellar
  Meteoroids}.
\newblock {\em \jgr}, 105:10353--10362.

\bibitem[{Begelman} and {Rees}, 1976]{BegelmanRees:1976}
{Begelman}, M.~C. and {Rees}, M.~J. (1976).
\newblock {Can Cosmic Clouds Cause Climatic Catastrophes}.
\newblock {\em \nat}, 261:298.

\bibitem[{Bochkarev}, 1984]{Bochkarev:1984}
{Bochkarev}, N.~G. (1984).
\newblock {Large-Scale Bubble Structure of the Interstellar Medium and
  Properties of the Local Spiral Arm}.
\newblock {In \emph{Local Interstellar Medium},}
\newblock {Eds. Y.~Kondo, F.~Bruhweiler, and B.~Savage}

\bibitem[{Bravo} et~al., 1998]{Bravoetal:1998}
{Bravo}, S., {Stewart}, G.~A., and {Blanco-Cano}, X. (1998).
\newblock {The Varying Multipolar Structure of the Sun's Magnetic Field and the
  Evolution of the Solar Magnetosphere Through the Solar Cycle}.
\newblock {\em \solphys}, 179:223--235.

\bibitem[{Burlaga} et~al., 2005]{Burlagaetal:2005}
{Burlaga}, L.~F., {Ness}, N.~F., {Acu{\~n}a}, M.~H., {Lepping}, R.~P.,
  {Connerney}, J.~E.~P., {Stone}, E.~C., and {McDonald}, F.~B. (2005).
\newblock {Crossing the Termination Shock into the Heliosheath: Magnetic
  Fields}.
\newblock {\em Science}, 309:2027--2029.

\bibitem[{Butler} et~al., 1978]{ButlerNewmanTalbot:1978}
{Butler}, D.~M., {Newman}, M.~J., and {Talbot}, R.~J. (1978).
\newblock {Interstellar Cloud Material - Contribution to Planetary
  Atmospheres}.
\newblock {\em Science}, 201:522--525.

\bibitem[{Caballero-Lopez} et~al., 2004]{CaballeroMcCracken:2004}
{Caballero-Lopez}, R.~A., {Moraal}, H., {McCracken}, K.~G., and {McDonald},
  F.~B. (2004).
\newblock {The Heliospheric Magnetic Field from 850 to 2000 AD Inferred from
  $^{10}$Be Records}.
\newblock {\em \jgr (Space Physics)}, 109:12102.

\bibitem[{Christl} et~al., 2004]{Christl:2004}
{Christl}, M., {Mangini}, A., {Holzk{\"a}mper}, S., and {Sp{\"o}tl}, C. (2004).
\newblock {Evidence for a link between the flux of galactic cosmic rays and
  Earth's climate during the past 200,000 years}.
\newblock { \em \jatp}, 66:313--322.

\bibitem[{Decker} et~al., 2005]{Deckeretal:2005}
{Decker}, R.~B., {Krimigis}, S.~M., {Roelof}, E.~C., {Hill}, M.~E.,
  {Armstrong}, T.~P., {Gloeckler}, G., {Hamilton}, D.~C., and {Lanzerotti},
  L.~J. (2005).
\newblock {Voyager 1 in the Foreshock, Termination Shock, and Heliosheath}.
\newblock {\em Science}, 309:2020--2024.

\bibitem[{Fahr}, 1968]{Fahr:1968}
{Fahr}, H.~J. (1968).
\newblock {On the Influence of Neutral Interstellar Matter on the Upper
  Atmosphere}.
\newblock {\em \apss}, 2:474.

\bibitem[{Fahr}, 2004]{Fahr:2004}
{Fahr}, H.-J. (2004).
\newblock {Global Structure of the Heliosphere and Interaction with the Local
  Interstellar Medium: Three Decades of Growing Knowledge}. \\
\newblock {\em \adsr}, 34:3--13.

\bibitem[Fahr et~al., 2006]{Fahretal:2006jos}
Fahr, Hans~J., Fichtner, Horst, Scherer, Klaus, and Stawicki, Olaf (2006).
\newblock {Variable Terrestrial Particle Environments During the Galactic Orbit
  of the Sun}.
\newblock In {\em {Solar Journey: The Significance of Our Galactic Environment
  for the Heliosphere and Earth, {in press}}}, page {000}. Springer.

\bibitem[Florinski and Zank, 2006]{FlorinskiZank:2006jos}
Florinski, Vladimir and Zank, Gary~P. (2006).
\newblock {The Galactic Cosmic Ray Intensity in the Heliosphere in Response to
  Variable Interstellar Environments}.
\newblock In {\em {Solar Journey: The Significance of Our Galactic Environment
  for the Heliosphere and Earth, {in press}}}, page {000}. Springer.

\bibitem[{Frisch} and {York}, 1986]{FrischYork:1986}
{Frisch}, P. and {York}, D.~G. (1986).
\newblock Interstellar clouds near the {S}un.
\newblock In {\em The {G}alaxy and the {S}olar {S}ystem}, pages 83--100.
  Eds. R.~Smoluchowski, J.~Bahcall, and M.~Matthews (University of Arizona Press).

\bibitem[{Frisch}, 1981]{Frisch:1981}
{Frisch}, P.~C. (1981).
\newblock The Nearby Interstellar Medium.
\newblock {\em \nat}, 293:377--379.

\bibitem[{Frisch}, 1993]{Frisch:1993a}
{Frisch}, P.~C. (1993).
\newblock G-star Astropauses - {A} Test for Interstellar Pressure.
\newblock {\em \apj}, 407:198--206.

\bibitem[Frisch, 1997]{Frisch:1997}
Frisch, P.~C. (1997).
\newblock Journey of the {S}un. 
\newblock {\em http://xxx.lanl.gov/}, astroph/9705231.

\bibitem[{Frisch}, 1998]{Frisch:1998}
{Frisch}, P.~C. (1998).
\newblock Interstellar Matter and the Boundary Conditions of the Heliosphere.
\newblock {\em \ssr}, 86:107--126.

\bibitem[Frisch, 2000]{Frisch:2000amsci}
Frisch, P.~C. (2000).
\newblock The {G}alactic Environment of the {S}un.
\newblock {\em American Scientist}, 88:52--59.

\bibitem[{Frisch}, 2005]{Frisch:2005L}
{Frisch}, P.~C. (2005).
\newblock {Tentative Identification of Interstellar Dust in the Magnetic Wall
  of the Heliosphere}.
\newblock {\em \apjl}, 632:L143--L146.

\bibitem[{Frisch} and {Slavin}, 2006]{FrischSlavin:2006jos}
{Frisch}, P.~C. and {Slavin}, J.~D. (2006).
\newblock {Short Term Variations in the Galactic Environment of the Sun}.
\newblock In {\em {Solar Journey: The Significance of Our Galactic Environment
  for the Heliosphere and Earth, {in press}}}, page {000}. Springer.

\bibitem[{Frisch} et~al., 1999]{Frischetal:1999}
{Frisch}, P.~C., {Dorschner}, J.~M., {Geiss}, J., {Greenberg}, J.~M., {Gr\"un},
  E., {Landgraf}, M., {Hoppe}, P., {Jones}, A.~P., {Kr{\"{a}}tschmer}, W.,
  {Linde}, T.~J., {Morfill}, G.~E., {Reach}, W., {Slavin}, J.~D., {Svestka},
  J., {Witt}, A.~N., and {Zank}, G.~P. (1999).
\newblock {Dust in the Local Interstellar Wind}.
\newblock {\em \apj}, 525:492--516.

\bibitem[{Gosling}, 1996]{Gosling:1996}
{Gosling}, J.~T. (1996).
\newblock {Corotating and Transient Solar Wind Flows in Three Dimensions}.
\newblock {\em \araa}, 34:35--74.

\bibitem[{Gruen} and {Landgraf}, 2000]{GruenLandgraf:2000}
{Gruen}, E. and {Landgraf}, M. (2000).
\newblock {Collisional Consequences of Big Interstellar Grains}.
\newblock {\em \jgr}, 105:10291--10298.

\bibitem[{Gurnett} and {Kurth}, 2005]{Gurnett:2005}
{Gurnett}, D.~A. and {Kurth}, W.~S. (2005).
\newblock {Electron Plasma Oscillations Upstream of the Solar Wind Termination
  Shock}.
\newblock {\em Science}, 309:2025--2027.

\bibitem[{Halley}, 1724]{Halley:1694}
{Halley}, E. (1724).
\newblock {Some Considerations about the Cause of the Universal Deluge, Laid
  before the Royal Society, on the 12th of December 1694. By Dr. Edmond Halley,
  R. S. S.}
\newblock {\em Philosophical Transactions Series I}, 33:118--123.

\bibitem[{Heiles} and {Troland}, 2003]{HTI}
{Heiles}, C. and {Troland}, T.~H. (2003).
\newblock {The Millennium Arecibo 21 Centimeter Absorption-Line Survey. I.
  Techniques and Gaussian Fits}.
\newblock {\em \apjs}, 145:329--354.

\bibitem[{Herschel}, 1795]{Herschel:1795}
{Herschel}, W. (1795).
\newblock {On the Nature and Construction of the Sun and Fixed Stars. By
  William Herschel, LL.D. F. R. S.}
\newblock {\em Philosophical Transactions Series I}, 85:46--72.

\bibitem[{Herschel}, 1801]{Herschel:1801}
{Herschel}, W. (1801).
\newblock {Observations Tending to Investigate the Nature of the Sun, in Order
  to Find the Causes or Symptoms of Its Variable Emission of Light and Heat;
  With Remarks on the Use That May Possibly Be Drawn from Solar Observations}.
\newblock {\em Philosophical Transactions Series I}, 91:265--318.

\bibitem[{Holzer}, 1989]{Holzer:1989}
{Holzer}, T.~E. (1989).
\newblock Interaction between the Solar Wind and the Interstellar Medium.
\newblock {\em \araa}, 27:199--234.

\bibitem[{Jokipii}, 2004]{Jokipii:2004}
{Jokipii}, J.~R. (2004).
\newblock {Transport of Cosmic Rays in the Heliosphere}.
\newblock In {\em AIP Conf. Proc. 719: Physics of the Outer Heliosphere}, pages
  249--259.

\bibitem[Kirkby and Carslaw, 2006]{KirkbyCarslaw:2006jos}
Kirkby, Jasper and Carslaw, Kenneth~S. (2006).
\newblock {Variations of Galactic Cosmic Rays and the Earth's Climate}.
\newblock In {\em {Solar Journey: The Significance of Our Galactic Environment
  for the Heliosphere and Earth, {in press}}}, page {000}. Springer.

\bibitem[{Lallement} et~al., 2005]{Lallementetal:2005}
{Lallement}, R., {Qu{\' e}merais}, E., {Bertaux}, J.~L., {Ferron}, S.,
  {Koutroumpa}, D., and {Pellinen}, R. (2005).
\newblock {Deflection of the Interstellar Neutral Hydrogen Flow Across the
  Heliospheric Interface}.
\newblock {\em Science}, 307:1447--1449.

\bibitem[{Landgraf}, 2000]{Landgraf:2000}
{Landgraf}, M. (2000).
\newblock {Modeling the Motion and Distribution of Interstellar Dust inside the
  Heliosphere}.
\newblock {\em \jgr}, 105:10303--10316.

\bibitem[Landgraf, 2006]{Landgraf:2006jos}
Landgraf, Markus (2006).
\newblock {Variations of the Interstellar Dust Distribution in the
  Heliosphere}.
\newblock In {\em {Solar Journey: The Significance of Our Galactic Environment
  for the Heliosphere and Earth, {in press}}}, page {000}. Springer.


\bibitem[{Langley}, 1876]{Langley:1876}
{Langley}, S.~P. (1876).
\newblock {Measurement of the Direct effect of Sun-Spots on Terrestrial
  Climates}.
\newblock {\em \mnras}, 37:5.

\bibitem[Linde, 1998]{Linde:1998}
Linde, T.~J. (1998).
\newblock {\em A Three-Dimensional Adaptive Multifluid {MHD} Model of the
  Heliosphere}.
\newblock PhD thesis, Univ. of Michigan, Ann Arbor. \\
\newblock {\emph{http://hpcc.engin.umich.edu/CFD/publications}}.

\bibitem[{Lockwood} and {Webber}, 2005]{LockwoodWebber:2005}
{Lockwood}, J.~A. and {Webber}, W.~R. (2005).
\newblock {Intensities of Galactic Cosmic Rays of 1.5 GeV Rigidity 
  versus Heliospheric Current Sheet Tilt}.
\newblock {\em \jgr (Space Physics)}, 110:4102.

\bibitem[{Lopate}, 2005]{Lopate:2005}
{Lopate}, C. (2005).
\newblock {Fifty Years of Ground Level Solar Particle Event Observations, In
  \emph{Solar Eruptions and Energetic Particles}}.
\newblock {\em \jgr}.

\bibitem[{M{\" o}bius} et~al., 2004]{Moebiusetal:2004}
{M{\" o}bius}, E., {Bzowski}, M., {Chalov}, S., {Fahr}, H.-J., {Gloeckler}, G.,
  {Izmodenov}, V., {Kallenbach}, R., {Lallement}, R., {McMullin}, D., {Noda},
  H., {Oka}, M., {Pauluhn}, A., {Raymond}, J., {Ruci{\' n}ski}, D., {Skoug},
  R., {Terasawa}, T., {Thompson}, W., {Vallerga}, J., {von Steiger}, R., and
  {Witte}, M. (2004).
\newblock {Synopsis of the Interstellar He Parameters from Combined Neutral
  Gas, Pickup Ion and UV Scattering Observations}.
\newblock {\em \aap}, 426:897--907.

\bibitem[M\"obius et~al., 2006]{Moebiusetal:2006jos}
M\"obius, Eberhard, Bzowski, Maciek, M\"uller, Hans-Reinhard, and Wurz, Peter
  (2006).
\newblock {Effects in the Inner Heliosphere Caused by Changing Conditions in
  the Galactic Environment}.
\newblock In {\em {Solar Journey: The Significance of Our Galactic Environment
  for the Heliosphere and Earth, {in press}}}, page {000}. Springer.


\bibitem[{McCracken} et~al., 2004]{McCrackenetal:2004}
{McCracken}, K.~G., {McDonald}, F.~B., {Beer}, J., {Raisbeck}, G., and {Yiou},
  F. (2004).
\newblock {A Phenomenological Study of the Long-Term Cosmic Ray Modulation,
  850-1958 AD}.
\newblock {\em \jgr (Space Physics)}, 109:12103.

\bibitem[{McCrea}, 1975]{McCrea:1975}
{McCrea}, W.~H. (1975).
\newblock {Ice Ages and the Galaxy}.
\newblock {\em \nat}, 255:607--609.

\bibitem[{McKay} and {Thomas}, 1978]{McKayThomas:1978}
{McKay}, C.~P. and {Thomas}, G.~E. (1978).
\newblock {Consequences of a Past Encounter of the Earth with an Interstellar
  Cloud}.
\newblock {\em \grl}, 5:215--218.

\bibitem[{Meisel} et~al., 2002]{Meiseletal:2002}
{Meisel}, D.~D., {Janches}, D., and {Mathews}, J.~D. (2002).
\newblock {Extrasolar Micrometeors Radiating from the Vicinity of the Local
  Interstellar Bubble}.
\newblock {\em \apj}, 567:323--341.

\bibitem[{Parker}, 1996]{Parker:1996}
{Parker}, E.~N. (1996).
\newblock {Solar Variability and Terrestrial Climate.}
\newblock In {\em The Sun and Beyond}, pages 117.

\bibitem[Parker, 2006]{Parker:2006jos}
Parker, Eugene~N. (2006).
\newblock {Interstellar Conditions and Planetary Magnetosphere}.
\newblock In {\em {Solar Journey: The Significance of Our Galactic Environment
  for the Heliosphere and Earth, {in press}}}, page {000}. Springer.

\bibitem[Pogorelov and Zank, 2006]{PogorelovZank:2006jos}
Pogorelov, Nikolai~V. and Zank, Gary~P. (2006).
\newblock {The Influence of the Interstellar Magnetic Field on the Heliospheric
  Interface}.
\newblock In {\em {Solar Journey: The Significance of Our Galactic Environment
  for the Heliosphere and Earth, {in press}}}, page {000}. Springer.


\bibitem[{Ratkiewicz} et~al., 1998]{Ratkiewicz:1998}
{Ratkiewicz}, R., {Barnes}, A., {Molvik}, G.~A., {Spreiter}, J.~R., {Stahara},
  S.~S., {Vinokur}, M., and {Venkateswaran}, S. (1998).
\newblock {Effect of Varying Strength and Orientation of Local Interstellar
  Magnetic Field on Configuration of Exterior Heliosphere}.
\newblock {\em \aap}, 335:363--369.

\bibitem[Richardson et~al., 1995]{Richardson:1995}
Richardson, J.~D., Paularena, K.~I., Lazarus, A.~J., and Belcher, J.~W. (1995).
\newblock {Radial Evolution of the Solar Wind from {IMP} 8 to {V}oyager 2}.
\newblock {\em \grl}, 22:325--328.

\bibitem[{Roble} and {Hays}, 1979]{RobleHaysII:1979}
{Roble}, R.~G. and {Hays}, P.~B. (1979).
\newblock {A Quasi-Static Model of Global Atmospheric Electricity. II -
  Electrical Coupling between the Upper and Lower Atmosphere}.
\newblock {\em \jgr}, 84:7247--7256.

\bibitem[{Scherer} and {Fahr}, 2003]{SchererFahr:2003}
{Scherer}, K. and {Fahr}, H.~J. (2003).
\newblock {Breathing of Heliospheric Structures Triggered by the Solar-cycle
  Activity}.
\newblock {\em Annales Geophysicae}, 21:1303--1313.

\bibitem[Shapley, 1921]{Shapley:1921}
Shapley, H. (1921).
\newblock Note on a Possible Factor in Changes of Geological Climate.
\newblock {\em J. Geology}, 29.

\bibitem[Shaviv, 2006]{Shaviv:2006jos}
Shaviv, Nir~J. (2006).
\newblock {Long-term Variations in the Galactic Environment of the Sun}.
\newblock In {\em {Solar Journey: The Significance of Our Galactic Environment
  for the Heliosphere and Earth, {in press}}}, page {000}. Springer.

\bibitem[{Simpson}, 2001]{Simpson:2001}
{Simpson}, J.~A. (2001).
\newblock {\em {The Cosmic Radiation}}, pages 117--152.
\newblock Century of Space Science, Volume I (Kluwer Academic Publishers).

\bibitem[{Smith} et~al., 2003]{Smithetal:2003}
{Smith}, E.~J., {Marsden}, R.~G., {Balogh}, A., {Gloeckler}, G., {Geiss}, J.,
  {McComas}, D.~J., {McKibben}, R.~B., {MacDowall}, R.~J., {Lanzerotti}, L.~J.,
  {Krupp}, N., {Krueger}, H., and {Landgraf}, M. (2003).
\newblock {The Sun and Heliosphere at Solar Maximum}.
\newblock {\em Science}, 302:1165--1169.

\bibitem[{Sonett}, 1992]{Sonett:1992}
{Sonett}, C.~P. (1992).
\newblock {A Supernova Shock Ensemble Model using Vostok $^{10}$Be Radioactivity}.
\newblock {\em Radiocarbon}, 34:239--245.

\bibitem[{Stone} et~al., 2005]{Stoneetal:2005}
{Stone}, E.~C., {Cummings}, A.~C., {McDonald}, F.~B., {Heikkila}, B.~C., {Lal},
  N., and {Webber}, W.~R. (2005).
\newblock {Voyager 1 Explores the Termination Shock Region and the Heliosheath
  Beyond}.
\newblock {\em Science}, 309:2017--2020.

\bibitem[{Talbot} and {Newman}, 1977]{TalbotNewman:1977}
{Talbot}, R.~J. and {Newman}, M.~J. (1977).
\newblock {Encounters Between Stars and Dense Interstellar Clouds}.
\newblock {\em \apjs}, 34:295--308.

\bibitem[{Thomas}, 1978]{Thomas:1978}
{Thomas}, G.~E. (1978).
\newblock {The Interstellar Wind and Its Influence on the Interplanetary
  Environment}.
\newblock {\em Annual Review of Earth and Planetary Sciences}, 6:173--204.

\bibitem[{Vall{\'e}e}, 2005]{ValleeSA:2005}
{Vall{\'e}e}, J.~P. (2005).
\newblock {The Spiral Arms and Interarm Separation of the Milky Way: An Updated
  Statistical Study}.
\newblock {\em \aj}, 130:569--575.

\bibitem[{Wang} and {Richardson}, 2005]{WangRichardson:2005}
{Wang}, C. and {Richardson}, J.~D. (2005).
\newblock {Dynamic Processes in the Outer Heliosphere: Voyager Observations and
  Models}.
\newblock {\em \adsr}, 35:2102--2105.

\bibitem[{Whang}, 2004]{Whang:2004}
{Whang}, Y.~C. (2004).
\newblock {Solar Cycle Variation of the Termination Shock}.
\newblock In {\em AIP Conf. Proc. 719: Physics of the Outer Heliosphere}, pages
  22--27.

\bibitem[{Willis}, 1978]{Willis:1978}
{Willis}, D.~M. (1978).
\newblock {Atmospheric Water Vapour of Extraterrestrial Origin - 
  Possible Role in Sun-Weather Relationships}.
\newblock {\em \jatp}, 40:513--528.

\bibitem[{Witte}, 2004]{Witte:2004}
{Witte}, M. (2004).
\newblock {Kinetic Parameters of Interstellar Neutral Helium. Review of Results
  obtained During One Solar Cycle with the Ulysses/GAS-Instrument}.
\newblock {\em \aap}, 426:835--844.

\bibitem[{Wood} et~al., 2005]{Woodetal:2005}
{Wood}, B.~E., {Redfield}, S., {Linsky}, J.~L., {M{\"u}ller}, H.-R., and
  {Zank}, G.~P. (2005).
\newblock {Stellar Ly{$\alpha$} Emission Lines in the Hubble Space Telescope
  Archive }.
\newblock {\em \apjs}, 159:118--140.

\bibitem[Yeghikyan and Fahr, 2006]{YeghikyanFahr:2006jos}
Yeghikyan, Ararat and Fahr, Hans~J. (2006).
\newblock {Accretion of Interstellar Material into the Heliosphere and onto
  Earth}.

\bibitem[{Zank} and {Frisch}, 1999]{ZankFrisch:1999}
{Zank}, G.~P. and {Frisch}, P.~C. (1999).
\newblock Consequences of a Change in the {G}alactic environment of the {S}un.
\newblock {\em \apj}, 518:965--973.

\bibitem[{Zank} and {M{\"u}ller}, 2003]{ZankMueller:2003}
{Zank}, G.~P. and {M{\"u}ller}, H.-R. (2003).
\newblock {The Dynamical Heliosphere}.
\newblock {\em \jgr (Space Physics)}, 108:7--1.

\bibitem[Zank et~al., 2006]{Zanketal:2006jos}
Zank, Gary~P., M\"uller, Hans-R., Florinski, Vladimir, and Frisch, Priscilla~C.
  (2006).
\newblock {Heliospheric Variation in Response to Changing Interstellar
  Environments}.
\newblock In {\em {Solar Journey: The Significance of Our Galactic Environment
  for the Heliosphere and Earth, {in press}}}, page {000}. Springer.


\end{chapthebibliography}

\emph{Submitted 15 November 2005; accepted 17 November 2005}
\end{document}